\documentclass[aps,reprint,twocolumn,superscriptaddress,citeautoscript,superscriptaddress,showpacs,amsmath,amssymb,amsfonts,floatfix,prb]{revtex4-1}

\usepackage{graphicx}
\usepackage{dcolumn}
\usepackage{bm}
\usepackage{times}
\usepackage{color}
\begin{document}
\title{Pressure-induced collapsed-tetragonal phase in SrCo$_{2}$As$_{2}$}

\author{W. T. Jayasekara}
\affiliation{Ames Laboratory, U.S. DOE, Iowa State University, Ames, Iowa 50011, USA}
\affiliation{Department of Physics and Astronomy, Iowa State University, Ames, Iowa 50011, USA}

\author{U. S. Kaluarachchi}
\affiliation{Ames Laboratory, U.S. DOE, Iowa State University, Ames, Iowa 50011, USA}
\affiliation{Department of Physics and Astronomy, Iowa State University, Ames, Iowa 50011, USA}

\author{B. G. Ueland}
\affiliation{Ames Laboratory, U.S. DOE, Iowa State University, Ames, Iowa 50011, USA}
\affiliation{Department of Physics and Astronomy, Iowa State University, Ames, Iowa 50011, USA}

\author{Abhishek Pandey}
\affiliation{Ames Laboratory, U.S. DOE, Iowa State University, Ames, Iowa 50011, USA}
\affiliation{Department of Physics and Astronomy, Iowa State University, Ames, Iowa 50011, USA}

\author{Y. B. Lee}
\affiliation{Ames Laboratory, U.S. DOE, Iowa State University, Ames, Iowa 50011, USA}
\affiliation{Department of Physics and Astronomy, Iowa State University, Ames, Iowa 50011, USA}

\author{V. Taufour}
\affiliation{Ames Laboratory, U.S. DOE, Iowa State University, Ames, Iowa 50011, USA}
\affiliation{Department of Physics and Astronomy, Iowa State University, Ames, Iowa 50011, USA}

\author{A. Sapkota}
\affiliation{Ames Laboratory, U.S. DOE, Iowa State University, Ames, Iowa 50011, USA}
\affiliation{Department of Physics and Astronomy, Iowa State University, Ames, Iowa 50011, USA}

\author{K. Kothapalli}
\affiliation{Ames Laboratory, U.S. DOE, Iowa State University, Ames, Iowa 50011, USA}
\affiliation{Department of Physics and Astronomy, Iowa State University, Ames, Iowa 50011, USA}

\author{N. S. Sangeetha}
\affiliation{Ames Laboratory, U.S. DOE, Iowa State University, Ames, Iowa 50011, USA}
\affiliation{Department of Physics and Astronomy, Iowa State University, Ames, Iowa 50011, USA}

\author{G. Fabbris}
\affiliation{Advanced Photon Source, Argonne National Laboratory, Argonne, Illinois 60439, USA}

\author{L. S. I. Veiga}
\affiliation{Advanced Photon Source, Argonne National Laboratory, Argonne, Illinois 60439, USA}

\author{Yejun Feng}
\affiliation{Advanced Photon Source, Argonne National Laboratory, Argonne, Illinois 60439, USA}

\author{A. M. dos Santos}
\affiliation{Quantum Condensed Matter Division, Oak Ridge National Laboratory, Oak Ridge, Tennessee 37831, USA}

\author{S. L. Bud'ko}
\affiliation{Ames Laboratory, U.S. DOE, Iowa State University, Ames, Iowa 50011, USA}
\affiliation{Department of Physics and Astronomy, Iowa State University, Ames, Iowa 50011, USA}

\author{B. N. Harmon}
\affiliation{Ames Laboratory, U.S. DOE, Iowa State University, Ames, Iowa 50011, USA}
\affiliation{Department of Physics and Astronomy, Iowa State University, Ames, Iowa 50011, USA}

\author{P. C. Canfield}
\affiliation{Ames Laboratory, U.S. DOE, Iowa State University, Ames, Iowa 50011, USA}
\affiliation{Department of Physics and Astronomy, Iowa State University, Ames, Iowa 50011, USA}

\author{D. C. Johnston}
\affiliation{Ames Laboratory, U.S. DOE, Iowa State University, Ames, Iowa 50011, USA}
\affiliation{Department of Physics and Astronomy, Iowa State University, Ames, Iowa 50011, USA}

\author{A. Kreyssig}
\affiliation{Ames Laboratory, U.S. DOE, Iowa State University, Ames, Iowa 50011, USA}
\affiliation{Department of Physics and Astronomy, Iowa State University, Ames, Iowa 50011, USA}

\author{A. I. Goldman}
\affiliation{Ames Laboratory, U.S. DOE, Iowa State University, Ames, Iowa 50011, USA}
\affiliation{Department of Physics and Astronomy, Iowa State University, Ames, Iowa 50011, USA}

\date{\today}

\begin{abstract}
We present high-energy x-ray diffraction data under applied pressures up to $p$ = 29 GPa, neutron diffraction measurements up to $p$ = 1.1 GPa, and electrical resistance measurements up to $p$~= 5.9 GPa, on SrCo$_2$As$_2$. Our x-ray diffraction data demonstrate that there is a first-order transition between the tetragonal (T) and collapsed-tetragonal (cT) phases, with an onset above approximately 6 GPa at $T$ = 7~K. The pressure for the onset of the cT phase and the range of coexistence between the T and cT phases appears to be nearly temperature independent. The compressibility along the \textbf{\emph{a}}-axis is the same for the T and cT phases whereas, along the \textbf{\emph{c}}-axis, the cT phase is significantly stiffer, which may be due to the formation of an As-As bond in the cT phase. Our resistivity measurements found no evidence of superconductivity in SrCo$_2$As$_2$ for $p\le$ 5.9~GPa and $T\ge$~1.8~K.  The resistivity data also show signatures consistent with a pressure-induced phase transition for $p$ $\gtrsim$ 5.5~GPa.  Single-crystal neutron diffraction measurements performed up to 1.1 GPa in the T phase found no evidence of stripe-type or A-type antiferromagnetic ordering down to 10~K.  Spin-polarized total-energy calculations demonstrate that the cT phase is the stable phase at high pressure with a $\frac{c}{a}$ ratio of 2.54. Furthermore, these calculations indicate that the cT phase of SrCo$_2$As$_2$ should manifest either A-type antiferromagnetic or ferromagnetic order.

\end{abstract}

\maketitle
\section{Introduction}

The body-centered tetragonal $A$Fe$_{2}$As$_{2}$ ($A$ = Ca, Sr, Ba) compounds have provided fertile ground for investigations of the interactions between lattice, magnetic and electronic degrees of freedom, and their impact upon unconventional superconductivity (SC) in the iron arsenide compounds \cite{Johnston_2010,P&G_2010,CanBud_2010,Stewart_2011,M&A_2010,Dai_2012,Dai_2015}. The superconducting ground state is realized upon the suppression of the magnetic phase transition, ubiquitous to the iron pnictides, through elemental substitutions or under applied pressure.  In Ba(Fe$_{1-x}$Co$_{x}$)$_2$As$_2$, for example, the substitution of only a few percent Co for Fe destabilizes the stripe-type antiferromagnetic (AFM) order by detuning the nesting condition between the electron and hole pockets,  allowing the superconducting ground state to appear in the presence of substantial magnetic fluctuations at the stripe-type AFM propagation vector, \textbf{Q}$_{\rm{stripe}}$ = ($\frac{1}{2}~\frac{1}{2}~1$).  Increased Co substitution ($x$~$>$~0.14) leads to a complete suppression of both stripe-type magnetic fluctuations \cite{Matan_2010} and SC\cite{Ni_2008}. These stripe-type magnetic fluctuations appear to be a key ingredient for SC in the $A$Fe$_{2}$As$_{2}$ family in particular, and the iron arsenides more generally.\cite{Johnston_2010,P&G_2010,CanBud_2010,Stewart_2011,M&A_2010,Dai_2012,Dai_2015}

Recent investigations of the end member $A$Co$_{2}$As$_{2}$ ($A$ = Ca, Sr, Ba) family have revealed interesting behavior which can provide new insight into the relationship between magnetism, structure, and SC in the iron arsenides.  At ambient pressure, CaCo$_{1.86}$As$_{2}$ crystallizes in the so-called collapsed tetragonal (cT) phase, which possesses the same tetragonal (T) ThCr$_{2}$Si$_{2}$ structure ($I4/mmm$) as SrCo$_{2}$As$_{2}$, BaCo$_{2}$As$_{2}$, and the parent $A$Fe$_{2}$As$_{2}$ compounds, but with a much reduced $\it c$ lattice parameter and unit cell volume. In the cT phase, CaCo$_{1.86}$As$_{2}$ manifests A-type AFM ordering [ferromagnetic (FM) \emph{\textbf{ab}} planes aligned antiferromagnetically along the \textbf{\emph{c}} axis] below $T_{\rm{N}}$ = 53-77 K,\cite{Cheng_2012,Ying_2012,Dante_2013,Vivek_2014} with the ordered moments lying along the \textbf{\emph{c}} axis.  This is a quite different behavior as compared to the closely related CaFe$_{2}$As$_{2}$ compound, which undergoes a first-order structural transition under applied pressure from the ambient pressure T or orthorhombic (O) structure (depending on the temperature) to a cT phase \cite{Milton_2008,Kreyssig_2008,Goldman_2009,Canfield_2009} in which the Fe magnetic moment is quenched.\cite{Pratt_2009,Gretarsson_2013,Soh_2013}

Neither SrCo$_{2}$As$_{2}$ nor BaCo$_{2}$As$_{2}$ exhibit long-range magnetic order down to $T$ = 1.8 K. Instead, both display an enhanced magnetic susceptibility that has been described as Stoner-enhanced paramagnetism close to a quantum instability.\cite{Sefat_2009,Pandey_2013} Nevertheless, our recent inelastic neutron scattering measurements \cite{Jayasekara_2013} on SrCo$_2$As$_2$ indicate that magnetic fluctuations occur at positions corresponding to \textbf{Q}$_{\rm{stripe}}$ = $\left(\frac{1}{2}\,  \frac{1}{2}\, 1\right)$, which is the same propagation vector that is found in the $A$Fe$_{2}$As$_{2}$ parent compounds and superconductors. Although ARPES measurements on both SrCo$_{2}$As$_{2}$ (Ref.~\onlinecite{Pandey_2013}) and BaCo$_{2}$As$_{2}$ (Refs.~\onlinecite{Xu_2013,Dhaka_2013}) did not reveal any obvious nesting features associated with \textbf{Q}$_{\rm{stripe}}$, as found for the $A$Fe$_{2}$As$_{2}$ compounds, density-functional theory calculations employing the local-density approximation have revealed maxima in the generalized susceptibility consistent with possibilities for stripe-type, A-type, or ferromagnetic ordering.\cite{Jayasekara_2013}  For the measured $\frac{c}{a}$ ratio of the lattice parameters in the T phase, calculations of the total energy suggest that the A-type AFM and FM ground states are nearly degenerate, but only slightly preferred over stripe-type AFM order.\cite{Jayasekara_2013}

Many issues regarding the origin of the stripe-type magnetic fluctuations, and their relationship to SC in the iron arsenides, remain unresolved. For example, it is not clear why the stripe-type magnetic fluctuations in $A$(Fe$_{1-x}$Co$_{x}$)$_2$As$_2$ systems are suppressed for intermediate values of $x$, only to reappear as $x~\rightarrow~1$ for, at least, SrCo$_{2}$As$_{2}$.\cite{Jayasekara_2013}  Whereas the absence of SC in the nonmagnetic cT phase of CaFe$_2$As$_2$ suggests that stripe-type magnetic fluctuations may be a necessary ingredient for SC in the iron arsenides, the absence of SC, down to at least 1.8~K, in the presence of such fluctuations in SrCo$_{2}$As$_{2}$ argues that they are not sufficient.  With respect to SrCo$_2$As$_2$ itself, given the near degeneracy of the different magnetic ground states from the total-energy calculations mentioned above, it is interesting to consider which magnetic ground state, or perhaps even SC, ultimately triumphs as one tunes the structure and magnetic interactions via either chemical substitution or applied pressure.

Since chemical substitutions can introduce disorder, impurity scattering effects, and localized strain, we have chosen to study SrCo$_2$As$_2$ under applied pressure.  Here, we describe high-energy x-ray diffraction measurements of SrCo$_2$As$_2$ at applied pressures up to $p$ = 29 GPa, neutron diffraction measurements up to $p$ = 1.1 GPa, and electrical resistance measurements up to $p$ = 5.9 GPa.  Our x-ray diffraction data at $T$ = 7~K show a first-order transition between the T and cT structures with an onset above approximately $p \gtrsim$ 6~GPa.  We find no evidence of SC in electrical resistivity measurements for pressures up to 5.9~GPa and temperatures down to 1.8~K. The resistivity data do, however, show signatures consistent with a pressure-induced phase transition for $p$ $\gtrsim$ 5.5~GPa. Our single-crystal neutron diffraction measurements performed up to 1.1 GPa found no evidence of stripe-type or A-type antiferromagnetic ordering down to 10~K in the T phase with an ordered moment greater than 0.4 $\mu_{\rm{B}}/$Co.  Using the structural information from our experimental study, we performed spin-polarized total-energy calculations to show that the cT phase is the lowest energy structure for the measured values of the volume change, $\Delta$$V$, with a resulting $\frac{c}{a}$ ratio of 2.5, close to the experimentally observed value.  In addition, these calculations indicate that SrCo$_2$As$_2$ should order into either an A-type AFM or FM structure in the cT phase.

\section{Experimental Details}
Single crystals of SrCo$_{2}$As$_{2}$ were grown from solution using Sn flux as described previously,\cite{Pandey_2013} and stored under inert gas.  Energy-dispersive x-ray (EDX) analysis using a JEOL-JSM-5910LV scanning electron microscope found no visible peaks associated with Sn incorporation into the flux-grown sample and the EDX software provided an upper limit on the Sn content of 0.028 at.-\%, consistent with previous findings.\cite{Pandey_2013}

Two sets of high-energy x-ray diffraction (HE-XRD) measurements were performed on station 6-ID-D at the Advanced Photon Source. In the first experiment, data were collected at $T$ = 7 K, from ambient pressure up to 29 GPa with an incident x-ray wavelength  $\lambda$ = 0.24204~\AA. A three-pin bronze alloy diamond anvil cell (DAC) was used with either silicon oil or a 4:1 mixture of methanol and ethanol as the pressure transmitting medium. In the second experiment, temperature dependent measurements were performed between $T$ = 7 and 300~K, for applied pressures ranging from $p$ = 1.8 GPa to 20 GPa using an incident x-ray wavelength  $\lambda$ = 0.12386~\AA. Copper-beryllium membrane-driven DACs were used and helium gas was loaded at 1 GPa to act as the pressure transmitting medium.\cite{Yejun_2014} In both experiments finely powdered samples were produced by carefully crushing single crystals, and then loaded into the DAC. Ruby spheres and gold powder were also loaded into the cells for calibrating the applied pressure.\cite{Chervin_2001, Fei_2007} The DAC was mounted on the cold finger of a He closed-cycle refrigerator with a base temperature of 7~K. The x-ray powder diffraction patterns were recorded using a MAR345 image plate detector positioned 874 mm and 1188 mm behind the sample in the first and second experiments, respectively. The resulting patterns were azimuthally integrated and calibrated using Si powder and cerium dioxide as standards. Data were collected in several runs in each experiment, with different loadings of diamond anvil cells.

Neutron diffraction measurements were performed using a 136 mg single crystal on the SNAP instrument at the Spallation Neutron Source. The single crystal was cut, oriented, and loaded into a NiCrAl alloy piston-cylinder pressure cell with the ($H$~$H$~$L$) reciprocal lattice plane coincident with the horizontal plane of the instrument. Using Fluorinert as the pressure medium and Pb powder as the pressure calibrant, we obtained a maximum pressure of $p$ = 1.1~GPa.

The electrical resistance of SrCo$_{2}$As$_{2}$ was measured by the standard four-probe method in a Quantum Design Inc., Physical Property Measurement System (PPMS) with the current applied in the \textbf{\emph{ab}} plane. Four Au wires (12.7\,$\mu$m diameter) were attached to the samples by spot welding. A modified Bridgman cell\cite{Colombier2007} was used with a 1:1 mixture of $n$-pentane:iso-pentane as the pressure medium. The solidification of this medium occurs in the range of $p$ $\sim$ 6 - 7\,GPa at ambient temperature\cite{Piermarini1973,Kim2011PRB}. The pressure was determined by the superconducting transition temperature of Pb,\cite{Bireckoven1988} measured by electrical resistance, and the solidification temperature of the medium is visible as a small anomaly in the temperature derivative of the resistivity of the sample.\cite{Kim2011PRB}

\section{Results and Discussion}
\subsection{High-energy X-ray Diffraction Measurements Under Applied Pressure}

\begin{figure}[!h]
	\centering
	\includegraphics[width=0.9\linewidth]{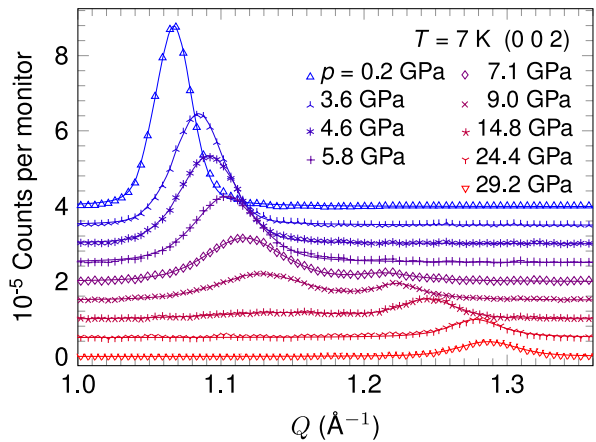}
	\caption{(Color online)  \label{Fig1} Evolution of the (0~0~2) HE-XRD powder Bragg peak as a function of applied pressure. The data at each pressure are offset by 0.5 $\times$ 10$^{-5}$ counts per monitor in the $\it y$-axis scale. The solid lines are fits to the data using Gaussian lineshapes.}

\end{figure}

\begin{figure}[!h]
	\centering
	\includegraphics[width=0.9\linewidth]{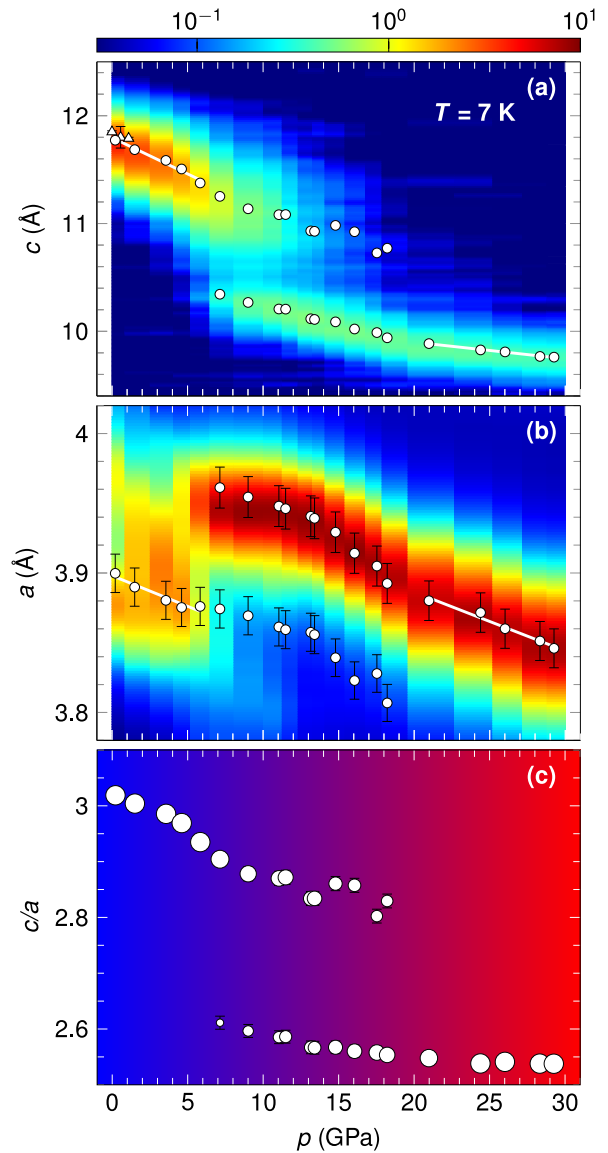}
	\caption{(color online)  \label{Fig2} Evolution of the unit cell dimensions with pressure. The detector counts are color-coded on a log scale in units of 10$^{-5}$ counts/monitor. (a) The $\it c$-lattice parameter determined from the (0 0 2) HE-XRD powder Bragg peak (solid white circles) and the (0 0 8) neutron single crystal Bragg peak (solid white triangles). (b) The $\it a$-lattice parameter determined from the (1 1 2) HE-XRD powder Bragg peak and the $c$-lattice parameter from panel (a). The solid white lines in (a) and (b) are linear fits to the pressure dependence of the lattice parameter. (c) The $\frac{c}{a}$ ratio determined from the data in panels (a) and (b).  The area of the circles represent the volume fractions of the T and cT phases and the background color provides a guide to the eye.}
\end{figure}

HE-XRD data were obtained from several runs using DAC configurations with different diamond culet sizes and pressure media.  Figure~\ref{Fig1} shows background-subtracted data in the vicinity of the (0~0~2) Bragg peak of SrCo$_2$As$_2$ taken at $T$ = 7~K using a DAC with a $500~\mu\rm{m}$ diameter culet, a $90~\mu\rm{m}$ thick pre-indented rhenium gasket with a $250~\mu\rm{m}$ diameter hole, and a 4:1 mixture of methanol and ethanol as the pressure medium.   Upon increasing pressure at $T$ = 7~K, Fig.~\ref{Fig1} shows that the (0~0~2) Bragg peak broadens and moves smoothly towards higher $Q$, where $Q$ = 2$\pi$/$d_{hkl}$, indicating a continuous decrease in the $c$-lattice parameter.  The peak broadening is likely due to non-hydrostatic pressure components present in the frozen pressure medium at low temperature. For $p$ $\gtrsim$ 6~GPa, a second well-separated peak appears at a higher $Q$, signalling the onset of the transition to the cT phase. This peak increases in intensity with increasing pressure as the lower-$Q$ peak, characteristic of the T phase,  diminishes and disappears above $\approx$ 18~GPa. These data demonstrate that there is a first-order phase transition between the T and cT phases with an extended pressure range of coexistence between 6 and 18 GPa as the volume fraction of the T phase decreases and the volume fraction of the cT phase increases. We note that the integrated intensity of the cT-phase (0~0~2) Bragg peak is smaller than that for the T structure.  This likely arises from either a decrease in the amount of sample illuminated by the beam, or a change in the degree of preferred orientation of the powder, as the pressure is increased.

The pressure evolution of the tetragonal unit cell dimensions at $T$ = 7~K was derived from fits to the (0~0~2) and (1~1~2) Bragg peaks and is summarized in Fig.~\ref{Fig2}. At ambient pressure, the $c$-lattice parameter of the T phase is 11.79(1) \AA~and decreases to 11.38(1) \AA~at $p$ = 5.8 GPa, where the first indication of the transition to the cT phase occurs.  There is a region of coexistence between the T and cT phases up to approximately 18 GPa in which a 7.9(3)\% reduction in the $\it c$-lattice parameter, a 2.1(3)\% increase in the $\it a$-lattice parameter, a 3.7(5)\% reduction in the unit cell volume, and a 9.9(5)\% decrease in the $\frac{c}{a}$ ratio occurs between both phases.

The in-plane (\textbf{\emph{a}}-axis), out-of-plane (\textbf{\emph{c}}-axis) and volume compressibilities were calculated from linear fits to the T and cT lattice parameters, excluding the region of coexistence, as shown by the solid white lines in Figs.~\ref{Fig2}(a) and \ref{Fig2}(b).  The results are listed in Table~\ref{tab:compress}. The compressibility along the \textbf{\emph{a}}-axis is the same for the T and cT phases, whereas along the \textbf{\emph{c}}-axis, the cT phase is significantly stiffer, which may be due to the formation of an As-As bond in the cT phase.\cite{Hoffmann_1985,Anand_2012}

\begin{table}
\caption{\label{tab:compress} Compressibilities of SrCo$_2$As$_2$ at $T$ = 7~K obtained from fits to the HE-XRD measurements as described in the text.}
\begin{ruledtabular}
\begin{tabular}{ccccccc}
  Structure & -$\frac{1}{a}\frac{\Delta\it{a}}{\Delta\it{p}}$ (GPa)$^{-1}$   &	 -$\frac{1}{c}\frac{\Delta\it{c}}{\Delta\it{p}}$ (GPa)$^{-1}$	&	-$\frac{1}{V}\frac{\Delta\it{V}}{\Delta\it{p}}$ (GPa)$^{-1}$\\	
\hline
    T & 0.0011(3)  	&	0.0057(5) 	&	0.0078(10)\\
   cT & 0.0011(3)	    &	0.0016(2) 	&	0.0037(7) \\
\end{tabular}
\end{ruledtabular}
\end{table}

\begin{figure}[!h]
	\centering
	\includegraphics[width=0.9\linewidth]{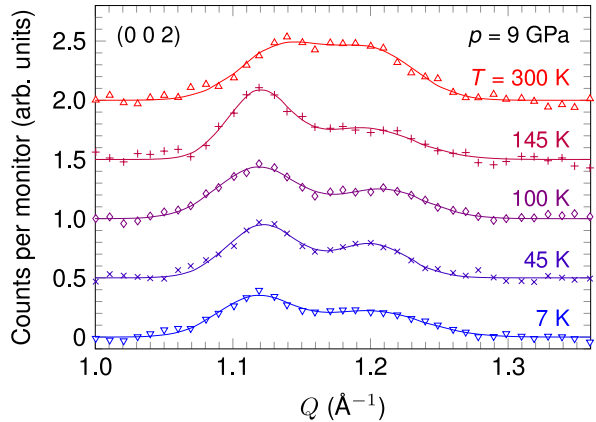}
	\caption{(color online)  \label{Fig3} The evolution of the (0 0 2) HE-XRD powder Bragg peak as a function of temperature at $p$ = 9~GPa. The data at each temperature are offset by 0.5 counts per monitor (arb. units) in the $\it y$-axis scale. The solid lines are fits to the data using Gaussian lineshapes.}
\end{figure}

We now turn to the second set of x-ray measurements on SrCo$_2$As$_2$ which focussed on the temperature dependence of the structural T - cT transition.   Figure~\ref{Fig3} shows the evolution of the (0 0 2) Bragg peak taken at different temperatures on cooling. Measurements were made using the membrane-driven Cu-Be DAC with a 600 $\mu$m diameter culet, a 50 $\mu$m thick pre-indented rhenium gasket with a 200 $\mu$m diameter hole, and helium gas as the pressure medium.  Upon increasing the pressure at ambient temperature the cT phase appears at $p$ $\gtrsim$~6.8 GPa, signaled by the presence of the higher-$Q$ peak above 1.2 \AA$^{-1}$.  As shown in Fig.~\ref{Fig3}, at $p~\simeq$ 9~GPa, both peaks remain in evidence and change only slightly as the temperature is lowered to our base temperature of 7~K, indicating that the T - cT transition $p$ - $T$ phase line is quite steep.

\subsection{Neutron Diffraction Measurements Under Applied Pressure}
To check for the possibility of A-type or stripe-type magnetic ordering under applied pressure up to $p$ = 1.1 GPa, neutron diffraction measurements were performed using a 136 mg single-crystal sample on the SNAP instrument at the Spallation Neutron Source as described in Section II. Fits to the (0~0~8) nuclear Bragg peak were used to determine the pressure dependence of the $c$-lattice parameter that is in good agreement with our x-ray data over the limited range in pressure [see Fig.~\ref{Fig2}(a)]. No evidence of either A-type or stripe-type magnetic Bragg scattering in the T phase was found down to $T$ = 10~K and up to $p$ = 1.1~GPa. Based on the measured intensity of several nuclear Bragg peaks and the background measured in the region of the expected magnetic peaks, we can place an upper limit of 0.4 $\mu_{\rm{B}}$/Co on the ordered moment for either A-type or stripe-type magnetic order, close to the estimated value of the moment for A-type AFM order in CaCo$_{1.86}$As$_2$.\cite{Cheng_2012,Dante_2013,Vivek_2014}

\subsection{Electrical Resistance Measurements Under Applied Pressure}
\begin{figure}[!h]
	\centering
	\includegraphics[width=1\linewidth]{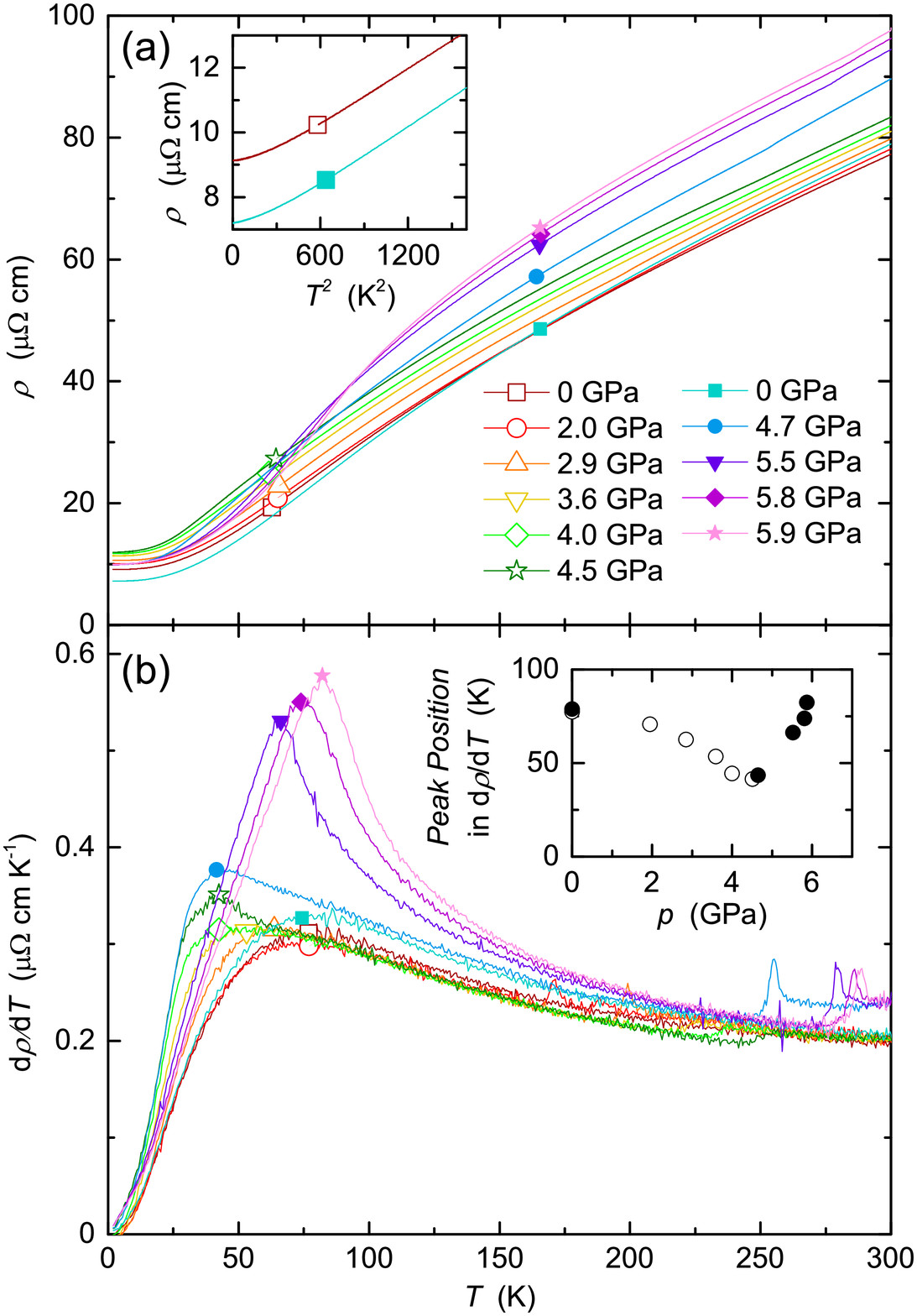}
	\caption{\label{Fig4} (color online) (a) Temperature dependence of the resistivity of SrCo$_{2}$As$_{2}$ up to $p$ = 5.9\,GPa obtained using a modified Bridgman cell.  The inset shows $\rho$($T$) vs. $T^2$ for $T \leq$~40~K. (b) The temperature derivative of the resistivity for several values of applied pressure. The sharp features above 200~K correspond to the freezing of the pressure medium. The inset shows the peak positions in d$\rho$/d$T$ as a function of pressure.  The open and filled symbols in all panels correspond to measurements on samples 1 and 2, respectively.  For $p$ $\geq$ 5.5~GPa, a sharp feature in the resistivity appears and moves up in temperature as $p$ increases.}	
	\end{figure}

The temperature-dependent resistance was measured on two samples under applied pressures up to $p$ = 5.9 GPa as shown in Fig.~\ref{Fig4}. The residual resistivity ratios (RRR) of the two samples were 8.5 (sample~1) and 10.9 (sample~2), which are somewhat smaller than the values previously measured (15.3),\cite{Pandey_2013} and the $T^2$ behavior of the resistivity reported in Ref.~\onlinecite{Pandey_2013} was not observed for these samples [see the inset to Fig.~\ref{Fig4}(a)].  The origin of the difference between the $T^2$ behavior observed in previous resistivity measurements and the present data is not yet clear.

Unlike several other FeAs-based compounds, such as  CaFe$_{2}$As$_{2}$\cite{Torikachvili2008PRL}, SrFe$_{2}$As$_{2}$ and BaFe${}_{2}$As${}_{2}$\cite{Colombier2009PRB,Kim2011PRB}, and KFe$_{2}$As$_{2}$ (Ref.~\onlinecite{Taufour2014PRB}) or another CoAs-based compound, BaCo$_{2}$As$_{2}$\cite{Ganguli2013}, we find that the room temperature resistivity of SrCo$_2$As$_2$ increases with increasing pressure. Between approximately $p$ = 2.0\,GPa and 4.5\,GPa we also observe a shallow upturn in the resistivity below $T$ = 5~K which prevents analysis of the pressure dependence of the temperature coefficient of the resistivity. Of importance here, however, is the absence of SC at all pressures measured for $T$ $\ge$ 1.8~K, and the presence of an anomaly in the resistivity below $\approx$ 100~K for $p$ $\gtrsim$ 5.5\,GPa. To examine this anomaly more closely, in Fig.~\ref{Fig4}(b) we plot the derivative of the resistivity as a function of temperature and the inset displays the peak position in d$\rho$/d$T$ as a function of pressure.  At ambient pressure, there is a broad, cross-over-like, peak at $\approx$ 77~K that moves down to 40~K with increasing pressure up to $\approx$ 4.7~GPa. Above $\approx$ 5.5~GPa, we find a distinct change in this feature that is likely related to the onset of the T to cT transition.  A sharp peak appears in d$\rho$/d$T$ and rapidly increases in temperature as pressure is increased.

\subsection{Spin-polarized Total Energy Calculations}
Using the results of our diffraction measurements, we performed spin-polarized calculations of the total energy to determine (1) if the cT phase is found as the stable structural phase at the experimentally determined volume reduction; and (2) how the preference for magnetic ordering is modified in the cT structure. We used the full potential linearized augmented plane wave (FPLAPW) method\cite{Blaha_2001} with a generalized gradient approximation functional (GGA),\cite{Perdew_1996} and employed 2.3, 2.0, 2.0 atomic unit muffin-tin radii for Sr, Co and As, respectively, with $R_{\rm{MT}}*k_{\rm{max}}$ = 8.0. Calculations were iterated, with 2400 \textbf{\emph{k}} points for the entire Brillouin zone, to reach the total energy convergence criterion of 0.01 mRy/cell. The experimentally determined unit cell dimensions were used, and the As positional coordinate $\it z$$_{\rm{As}}$ = 0.3588 (Ref.~\onlinecite{Pandey_2013}) was held constant for all calculations.

\begin{figure}[!h]
	\centering
	\includegraphics[width=0.9\linewidth]{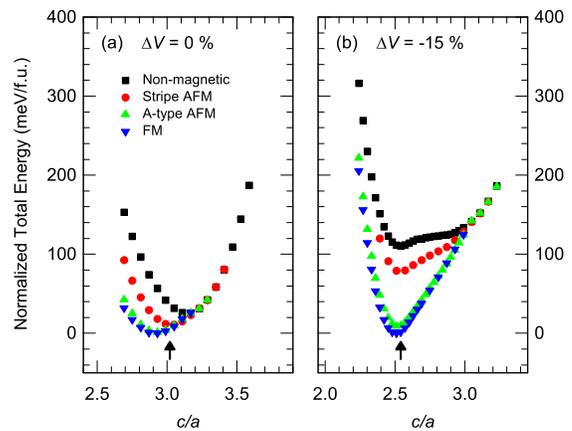}
	\caption{\label{Fig5} (color online) The energy difference between magnetic states from total energy calculations for (a) the T phase ($\Delta$$V$= 0\%) and (b) the cT phase ($\Delta V = -15$\%). The measured values of $\frac{c}{a}$ at ambient pressure and above 20~GPa are indicated by the black arrows.}
    \end{figure}

Figure~\ref{Fig5} shows the results for two different unit cell volumes corresponding to volume changes of $\Delta$$V$= 0\% (T structure) and $\Delta$$V$= $-15$\% (cT structure for $p \approx$ 20\,GPa).  The total energy calculations at ambient pressure are consistent with our previous calculations.\cite{Jayasekara_2013} The total-energy calculations for $\Delta$$V$= $-15$\% show that the cT phase is the minimum energy structure with a value of $\frac{c}{a}$ = 2.54, close to the experimentally observed value of 2.55 at $p$ = 21~GPa.

We also find that both the FM and A-type AFM order minimize the total energy and are nearly degenerate, whereas the total energy for stripe-type magnetic order lies somewhat above these values. We note that even in the T structure, magnetic order is predicted by these calculations for the observed value of $\frac{c}{a}$ indicated by the arrows in Fig.~\ref{Fig5}. We speculate that the near degeneracy in the magnetic ground state energies for the FM and A-type AFM order introduces some degree of frustration, suppressing long-range magnetic order at ambient pressure.  Our calculations still evidence a near-degeneracy between FM and A-type AFM order in the cT phase although, now, the stripe-type magnetic order lies higher in energy than either of these alternatives.  These results suggest that magnetic ordering in the cT phase should be either A-type AFM or FM, despite the presence of strong stripe-type magnetic fluctuations at ambient pressure.\cite{Jayasekara_2013}

\section{Summary}
We have identified the onset of a pressure-induced cT phase for SrCo$_{2}$As$_{2}$ at $p$ $\gtrsim$ 6 GPa for $T$ = 7~K. The transition between the T and cT phases appears to be first-order with an extended region of phase coexistence. Beyond 18 GPa, only the cT phase is observed.  Our x-ray and resistivity data also indicate that the T - cT transition $p$ - $T$ phase line is quite steep. Down to 1.8~K, our electrical resistance measurements find no evidence of superconductivity up to our maximum pressure of 5.9~GPa but an anomalous change in d$\rho$/d$T$, likely associated with the T - cT transition, appears above $p$ = 5.5~GPa. The peak position of this feature rapidly increases in temperature with pressure. The compressibilities of the T and cT phases along the \textbf{\emph{a}}-axis are the same, within error, but differ strongly along the \textbf{\emph{c}}-axis, as the cT phase is more than a factor of three stiffer.  Neutron diffraction measurements up to 1.1 GPa failed to identify A-type or stripe-type magnetic order for applied pressures $p \leq 1.1$~GPa and temperatures $T \geq 10$~K with an upper limit of 0.4 $\mu_{\rm{B}}$/Co.  Our total-energy calculations confirm that the cT phase, with a $\frac{c}{a}$ ratio of 2.54, is the stable structural phase at high pressure and suggest that the magnetic ordering in the cT phase should be either A-type AFM or FM.  Further magnetic neutron diffraction measurements on single crystal samples at higher pressures and x-ray magnetic circular dichroism measurements are planned to search for possible magnetic order in the cT phase.

\begin{acknowledgments}
The authors gratefully acknowledge the assistance of D. Robinson, B. Lavina, S. Tkachev, C. Kenney-Benson and S. Sinogeiken with the HE-XRD measurements and useful discussions with  D. Haskel and J. C. Lang. Work at the Ames Laboratory was supported by the Department of Energy, Basic Energy Sciences, Division of Materials Sciences \& Engineering, under Contract No. DE-AC02-07CH11358.  This research used resources of the Advanced Photon Source, a U.S. Department of Energy (DOE) Office of Science User Facility operated for the DOE Office of Science by Argonne National Laboratory under Contract No. DE-AC02-06CH11357. Research conducted at the ORNL Spallation Neutron Source was sponsored by the Scientific User Facilities Division, Office of Basic Energy Sciences, U.S. Department of Energy. Use of the COMPRES-GSECARS gas loading system was supported by COMPRES under NSF Cooperative Agreement EAR 11-57758 and by GSECARS through NSF Grant EAR-1128799 and DOE Grant DE-FG02-94ER14466. Portions of this work were performed at HPCAT (Sector 16), Advanced Photon Source (APS), Argonne National Laboratory. HPCAT operations are supported by DOE-NNSA under Award No. DE-NA0001974 and DOE-BES under Award No. DE-FG02-99ER45775, with partial instrumentation funding by NSF.
\end{acknowledgments}

\end{document}